\documentclass[aps,pre,
onecolumn,
twocolumn,
10pt,
notitlepage,
showpacs,floatfix,nofootinbib,
superscriptaddress,
]{revtex4-1}

\usepackage{subfigure}
\usepackage{amssymb}
\usepackage{amsfonts}
\usepackage{amsmath}
\usepackage{amsthm}
\usepackage{epsfig}
\usepackage{float}
\usepackage{multirow}
\usepackage[usenames,dvipsnames]{color}
\usepackage[latin1]{inputenc}

\usepackage{enumitem} 

\usepackage{graphics,graphicx,xcolor,subfigure}

\usepackage{xr-hyper}
\usepackage[hidelinks,unicode=true]{hyperref}
\hypersetup{colorlinks=true,% Don't draw a box around links, instead color every piece of text, which is a link
	linkcolor=blue,
	urlcolor=blue,
	citecolor=blue,
	filecolor=blue,
	pdfhighlight=/N% When clicking a link and holding the mouse button, don't use any graphical effects - i.e. the alternative would be to behave like a "button" which is pressed within the PDF
}%\usepackage{hyperref}

\usepackage{comment}
\usepackage[normalem]{ulem}

\newcommand{\lbA}{\lambda_\text{A}}
\newcommand{\lbI}{\lambda_\text{I}}
\newcommand{\muA}{\mu_\text{A}}
\newcommand{\betaR}{\beta_\text{R}}
\newcommand{\betaI}{\beta_\text{I}}
\newcommand{\betaC}{\beta_\text{C}}
\newcommand{\alphaR}{\alpha_\text{R}}
\newcommand{\alphaC}{\alpha_\text{C}}
\newcommand{\gammaA}{\gamma_\text{A}}
\newcommand{\gammaI}{\gamma_\text{I}}
\newcommand{\CA}{C_\text{A}}
\newcommand{\CI}{C_\text{I}}
\newcommand{\RA}{R_\text{A}}
\newcommand{\RI}{R_\text{I}}
\newcommand{\pA}{p_\text{A}}
\newcommand{\pI}{p_\text{I}}

\newcommand{\orcid}[1]{\hspace{0.2em}\href{https://orcid.org/#1}{\includegraphics[keepaspectratio,width=0.7em]{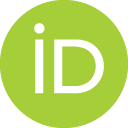}}}

\begin{document}
	
	\title{
		%	Estimating undocumented infections and role of nonpharmacological interventions  using reported cases time series: Application to SARS-CoV-2 in Brazil.\\or\\  
		%	
		Data-driven approach in a compartmental epidemic model to assess undocumented infections}
	
	\author{Guilherme S. Costa\orcid{0000-0002-5019-0098}}
	\affiliation{Departamento de F\'{\i}sica, Universidade Federal de Vi\c{c}osa, 36570-900 Vi\c{c}osa, Minas Gerais, Brazil}

	\author{Wesley Cota\orcid{0000-0002-8582-1531}}
	\affiliation{Departamento de F\'{\i}sica, Universidade Federal de Vi\c{c}osa, 36570-900 Vi\c{c}osa, Minas Gerais, Brazil}
	
	\author{Silvio C. Ferreira\orcid{0000-0001-7159-2769}}
	%\email{silviojr@ufv.br}
	%\thanks{Corresponding author.}
	\affiliation{Departamento de F\'{\i}sica, Universidade Federal de Vi\c{c}osa, 36570-900 Vi\c{c}osa, Minas Gerais, Brazil}
	\affiliation{National Institute of Science and Technology for Complex Systems, 22290-180, Rio de Janeiro, Brazil}

	\begin{abstract}
		Nowcasting and forecasting of epidemic spreading rely on incidence series of reported cases to derive the fundamental epidemiological parameters for a given pathogen. Two relevant drawbacks for predictions are the unknown fractions of undocumented cases and levels of nonpharmacological interventions, which span highly heterogeneously across different places and times. We describe a simple data-driven approach using a compartmental model including asymptomatic and pre-symptomatic contagions that allows to estimate both the level of undocumented infections and the value of effective reproductive number $R_t$ from time series of reported cases, deaths, and epidemiological parameters. The method was applied to epidemic series for COVID-19 across different municipalities in Brazil allowing to estimate the heterogeneity level of under-reporting across different places. The reproductive number derived within the current framework is little sensitive to both diagnosis and infection rates during the asymptomatic states. The methods described here can be extended to more general cases if data is available and adapted to other epidemiological approaches and surveillance data.
	\end{abstract}
	%\pacs{}
	
	\maketitle

\section{Introduction}

Our contemporary society is facing an unprecedented threat imposed by the COVID-19, caused by the pathogen SARS-CoV-2, evidencing the importance, limitations, and subtleties of using compartmental epidemic models for the forecasting or nowcasting of pandemic scenarios~\cite{Wu2020,Zhang2020,Estrada2020,Gilbert2020,Li2020}. After two years of intensive investigation,  much has been learned with respect to the virology of SARS-CoV-2 in humans~\cite{Li2020b,Nishiura2020,Baud2020,He2020}. Among other achievements, several key aspects of the transmission were unveiled~\cite{Tindale2020, Li2020, Emery2020, Davies2020} and efficient vaccines have been developed~\cite{Lipsitch2020}. Variants of {of the original strain}~\cite{Davies2021,Buss2021} give rise to new and more aggressive outbreaks due to reinfection and raised contagion rates that tend to become endemic, circulating among humans indefinitely with new outbreaks emerging seasonally~\cite{Sridhar2021}. Whilst the biology of the virus and interaction with human hosts is better understood, other crucial  aspects of the  epidemiology, specially the behavioral ones, remains unpredictable even at a short-term, varying across time and location. In particular, the non-pharmaceutical interventions (NPIs), such as face masks, testing policies and social distancing have played a central role on the spreading of SARS-CoV-2~\cite{Pan2020,Ali2020,Goyal2021,Flaxman2020}. The aforementioned NPIs  contribute for reduction of the contagion rates in an uncontrolled way, such that the effective contagion rate must be inferred along the time from count case series via likelihood or other calibration methods~\cite{Cori2013,Gostic2020}.  

A fundamental epidemic characteristic of the SARS-CoV-2 contagion in humans is its high transmission before the onset of the symptoms~\cite{Li2020,Emery2020,Mizumoto2020}, the presymptomatic individuals, and even the contagion by those who never manifest relevant symptoms~\cite{Byambasuren2020,Buitrago-Garcia2020}, the true asymptomatic individuals. The latter could be accessed by mass testing and contact tracing, for example. Seroprevalence studies for different  phases and regions~\cite{Davies2021,Ioannidis2021} reveal  population incidences of antibodies for SARS-CoV-2 in levels much higher than the {case} counts reported {by the epidemiological}  surveillance systems. So, the case fatality ratio (CFR), defined as the ratio between the {numbers} of diagnosed deaths and  cases, can differ substantially from the infection fatality ratio (IFR), defined as the fraction of all infections (documented or not) that evolve to death~\cite{Ioannidis2021,Verity2020}.

The level of under-reporting, in which the CFR is greater than the IFR, varies widely in different seroprevalence inquiries~\cite{Ioannidis2021} due to several uncontrolled factors such as the testing policies (only symptomatic cases, contact tracing, etc.), availability of tests (low or high income places), sensitivity of tests (antigen or PCR),  and  seeking for medical care, among others. The relation between seropositivity and immunity is not fully established and new emerging variants always open paths for reinfections and new outbreaks~\cite{Romano2021}. Therefore, to estimate the level of undocumented infections across different places and times remains a challenge. Epidemic models of statistical inference were developed to access the amount of undocumented infections of SARS-CoV-2. For example, Pullano et al.~\cite{Pullano2021} estimated that 9 out 10 cases of symptomatic infections were not ascertained by the surveillance system in France from 11 May to 28 June 2020, during the first epidemic wave of COVID-19, suggesting that large numbers of symptomatic cases of COVID-19 did not seek for medical advice. Lu et al.~\cite{Lu2021} considered four complementary approaches to estimate the cumulative incidence of symptomatic cases of COVID-19 in the US and concluded that on April 4 of 2020 the estimated case count was 5 to 50 times higher than the official counts of positive tests across the different states. Subramanianan et al.~\cite{Subramanian2021} used a model including testing information to fit the case and serology data from New York City, from March to June of 2020, to estimate a low proportion of symptomatic cases (13 to 18\% of the total infections), and that the reproductive number could be larger than often assumed. Similarly,  Irons and Raftery~\cite{Irons2021} used a similar approach to estimate that approximately 60\% of the infections were  not diagnosed by tests in USA as of March 7, 2021. Hallal et al.~\cite{Hallal2020} carried out two seroprevalence studies, the first in May 2020 and the second in June 2020,  in 133 municipalities of Brazil and estimated that only 10.3\% of all infections were documented.

Due to the importance of asymptomatic or pre-symptomatic transmission, the corresponding compartments were soon included in mathematical models for COVID-19~\cite{Emery2020,Costa2020,Arenas2020,Ashcroft2021,Ferretti2020}. However, it is concomitantly an additional source of uncertainty in the initial conditions. Predictive scenarios of the first SARS-CoV-2 outbreak were either semi-quantitative~\cite{Costa2020,Aleta2020,Aleta2020b} or based on Bayesian inference using reported cases' series~\cite{Arenas2020,Dehning2020,Maier2020}. Brazil is an  example, certainly not an exception, of highly heterogeneous responses to COVID-19 pandemics due to the lack of coordinated policies across different administrative layers~\cite{Castro2021}, in addition to the intrinsic variability of social-economic indexes across the country impacting directly the epidemiological outcomes. Therefore, a mechanistic approach for simulation of epidemic spreading with asymptomatic transmission calls for a systematic way to determine the initial conditions. 

The contribution of asymptomatic infections and testing policies to the effective reproductive number $R_t$~\cite{keeling2008modeling} through surveillance counts is an important issue~\cite{Irons2021,Subramanian2021}. The basic reproductive number  is defined as the average number of secondary infections generated by a single infected individual introduced in a completely susceptible population, commonly represented by $R_0$. The effective reproductive number is given by $R_t=S(t)R_0$, where $S(t)$ is the fraction of susceptible population (who can be infected by the pathogen) at time $t$. This definition, under the hypothesis of homogeneous mixing, is the simplest one and can be generalized to stratified compartments~\cite{keeling2008modeling}. The reproductive number can also be estimated directly from case counts using statistical inference models~\cite{Cori2013}, as reported for COVID-19 pandemics across the world~\cite{Ali2020,Arenas2020,Castro2021,Ferguson2020}.

In this present work, we describe a mechanistic approach to estimate the number of undocumented {infections} (symptomatic or not) using the epidemic surveillance data for confirmed cases and deaths. The method is grounded on a compartmental epidemic model including both documented and undocumented compartments, the latter not counted by epidemiological surveillance. The present approach allows to determine the effective reproductive number, the level of under-reporting and initial conditions using the date of diagnosis. The approach can be promptly modified or generalized for other types of data, epidemic compartments, and population stratification. The method shares similarities with the recent approaches to estimate undocumented cases~\cite{Pullano2021,Lu2021,Byambasuren2020,Irons2021}, such as the use of reported infections and deaths. The central difference is that our approach is essentially  mechanistic {and not Bayesian}.

We applied the method across different geographical scales of two Brazilian states, namely Paraná (PR) and Espírito Santo (ES), using time series  with  dates of COVID-19 diagnosis  available by the epidemiological surveillance of the respective states. The time window of investigation was from 1 January to 31 July of 2021, encompassing the second epidemic wave driven mainly by the Gamma variant~\cite{Castro2021,Sabino2021}. We report variable levels of under-reporting across different places and times. We were able to estimate initial conditions for the hidden compartments and effective infection rates along the time, which gave an efficient short-time forecast for the series of confirmed cases. Despite the basic reproductive number being explicitly dependent on the asymptomatic transmission, the analysis indicates that undocumented infections seem to not alter significantly the effective reproductive number for the analyzed series. 

The remaining of this paper is organized as follows. The methodology is detailed in section~\ref{sec:model}. The epidemic compartmental model and some analytical results are presented in subsection~\ref{sub:model}. The  data-driven approach to estimate the under-reporting level from {epidemiological} surveillance counts is described in subsection~\ref{sub:ur} while the eigenvalue approach to determine the initial conditions is presented in subsection~\ref{sub:hidden}. Application of the method to epidemiological data is presented in section~\ref{sec:results} and the main conclusions of the work are discussed in section~\ref{sec:conclu}. 

\section{A mechanistic approach to estimate undocumented cases}
\label{sec:model}

\subsection{Compartmental model}
\label{sub:model}
\begin{figure*}[!t]
	\centering
	\includegraphics[width=0.6\linewidth]{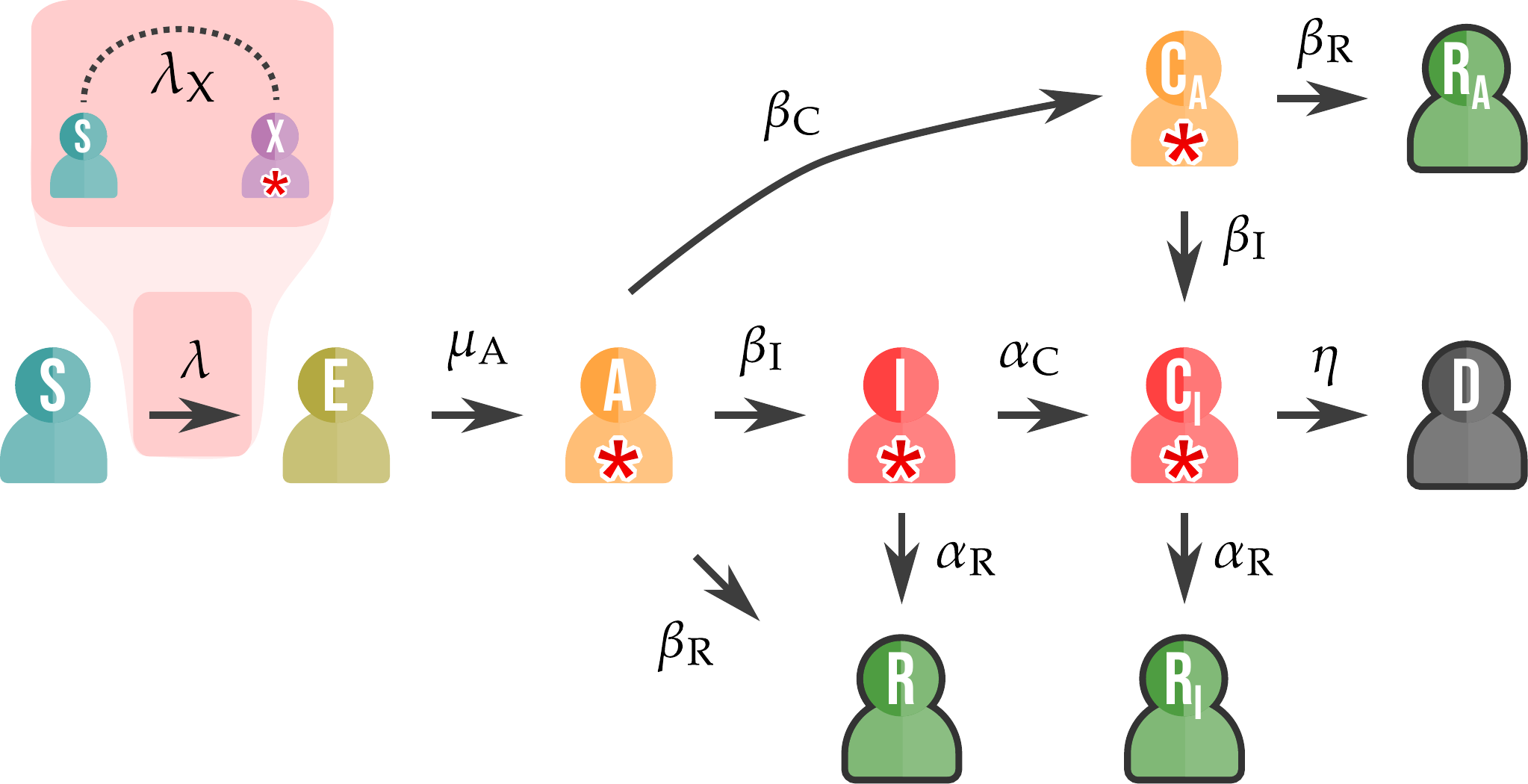}
	\caption{Schematic representation of the  epidemic model  including the following compartments: susceptible (S), exposed (E), asymptomatic (A), symptomatic (I), recovered (R, R$_\text{A}$, and R$_\text{I}$), deceased (D), and confirmed cases (C$_\text{A}$ and C$_\text{I}$). The transition and respective rates are indicated by arrows. The infectious compartments are depicted with the  symbol $\star$. The infection processes, represented by the dashed line, involve the interaction between susceptible and one of the infectious compartments, happening with rates $\lambda_\text{X}$, X$=$A, I, C$_\text{A}$, and C$_\text{I}$,  which {may} depend on the compartment.}
	\label{fig:modelo}
\end{figure*}
Following a mechanistic approach for population fractions, an epidemic process with presymptomatic, asymptomatic, and undocumented transmissions are investigated using a  compartmental model~\cite{keeling2008modeling}  under the  homogeneous mixing hypothesis. Individuals are grouped according to their epidemic states in the following compartments: \textit{Susceptible} (S) who can be infected; \textit{exposed} (E) who were infected but is not contagious yet; \textit{asymptomatic} (A) who are infectious but do not present symptoms; \textit{symptomatic} (I) ones who may seek for medical care due to the presence of symptoms; \textit{undocumented recovered} (R) who have been infected, healed but not diagnosed; \textit{deceased} (D) who died due to COVID-19;  two  compartments of \textit{diagnosed cases} for SARS-CoV-2 including individuals who were \textit{asymptomatic} (C$_\text{A}$) or \textit{symptomatic} (C$_\text{I}$) at {the} moment of testing; and the corresponding \textit{recovered compartments for confirmed cases} R$_\text{A}$ and R$_\text{I}$. {We assume constant rates and spontaneous transitions implying that the time last in a given infectious compartment is exponentially distributed~\cite{keeling2008modeling},  in contrast with the Biology of infectious pathogens where one expects a peaked distribution that excludes very short {and very long} exiting times. However, multiple infectious compartments soften the problem producing peaked distributions for the total infectious time with a negligible probability of recovering shortly~\cite{keeling2008modeling,Alexei2021}. The epidemiological model and rates are schematically depicted in Fig.~\ref{fig:modelo}. 

Susceptible persons in contact with infectious  individuals {(asymptomatic or symptomatic)}  become exposed with rates $\lbA$ and $\lbI$, respectively.  For sake of simplicity, confirmed cases are assumed to be isolated and do not contribute for new infections. The remaining transitions are represented in Fig.~\ref{fig:modelo}. Exposed individuals become asymptomatic with rate $\muA$. The latter can evolve to a symptomatic state with rate $\betaI$, recover  with rate $\betaR$, or be diagnosed by tests with rate $\betaC$ moving to the confirmed compartment C$_\text{A}$.  Similarly, the undocumented symptomatic individuals can recover with rate  $\alphaR$ or  be diagnosed and become C$_\text{I}$ with rate $\alphaC$.  The clinical state of confirmed cases evolves as does the undocumented ones. A confirmed case (C$_\text{I}$) can die (D) with rate $\eta$ while undocumented deaths are neglected, again, for sake of simplicity. The true asymptomatic and the presymptomatic cases are implicitly considered with transitions A$\rightarrow$R (C\textsubscript{A}$\rightarrow$R\textsubscript{A}) and  A$\rightarrow$I (A$\rightarrow$C\textsubscript{A}$\rightarrow$C\textsubscript{I}), respectively. {Compartments C\textsubscript{A} and C\textsubscript{I}  are simplifications of a more complex dynamics including seeking for test, {time for }results, and isolation.}

Assuming a constant population $N=\sum_\text{X} N_\text{X}$, where $N_\text{X}$ is the number of individuals in the compartment X, the above transitions can be summarized in the following set of differential equations
\begin{subequations}
	\begingroup
	\allowdisplaybreaks
	\begin{equation}
	%\frac{\dd S}{\dd t} 
	\partial_t{S}= -(\lbA A+\lbI I)S\,, \label{eq:dSdt}
	\end{equation}
	\begin{equation}
	%\frac{\dd E}{\dd t} 
	\partial_t{E} = (\lbA A+\lbI I)S -\muA E\,, \label{eq:dEdt} 
	\end{equation}
	\begin{equation}
	%\frac{\dd A}{\dd t} 
	\partial_t{A} = \muA E -(\betaI+\betaR+\betaC)A\,, \label{eq:dAdt} 
	\end{equation}
	\begin{equation}
	%\frac{\dd I}{\dd t} 
	\partial_t{I} = \betaI A-(\alphaR+\alphaC)I\,, \label{eq:dIdt}
	\end{equation}
	\begin{equation}
	%\frac{\dd R}{\dd t} 
	\partial_t{R} = \alphaR I+\betaR A\,, \label{eq:dRdt}
	\end{equation}
	\begin{equation}
	%\frac{\dd C}{\dd t} 
	\partial_t{C}= \betaC A +\alphaC I\,, \label{eq:dCdt}
	\end{equation}
	\begin{equation}
	%\frac{\dd \CA}{\dd t} 
	\partial_t{\CA}= \betaC A -(\betaR+\betaI)\CA\,,
	\end{equation}
	\begin{equation}
	%\frac{\dd \CI}{\dd t} 
	\partial_t{\CI}= \alphaC I +\betaI\CA -(\alphaR+\eta)\CI\,,
	\end{equation}
	\begin{equation}
	%\frac{\dd \RA}{\dd t}  
	\partial_t{\RA}= \betaR \CA\,,
	\end{equation}
	\begin{equation}
	%\frac{\dd \RI}{\dd t}  
	\partial_t{\RI}= \alphaR \CI\,,
	\end{equation}
	\begin{equation}
	%\frac{\dd D}{\dd t}  
	\partial_t{D}= \eta\CI\,,
	\end{equation}
	\endgroup
\end{subequations}
where $X=N_\text{X}/N$, $\text{X} \in\lbrace \text{S, E,\ldots, D}\rbrace$, is the corresponding population fraction in the compartment X.

The basic reproductive number $R_0$ is straightforwardly computed and given by
\begin{equation}
R_0 = \frac{1}{\betaI+\betaC+\betaR}\left[\lbA+
\lbI \frac{\betaI}{\alphaC+\alphaR}\right]\,.
\end{equation}

Consider a more intuitive parameterization in terms of the probabilities $\pA$ and $\pI$ that infected individuals are diagnosed during the asymptomatic or symptomatic phases, respectively, which can be computed from the compartmental model and are given by 
\begin{equation}
\pA=\frac{\betaC}{\betaI+\betaC+\betaR} \qquad \text{and} \qquad \pI=\frac{\alphaC}{\alphaC+\alphaR}\,.
\end{equation}
One can also show that an exposed individual ends diagnosed with probability
\begin{equation}
\mathcal{P}_\text{C} = \pA+(1-\pA)\phi\pI\,,
\label{eq:PC}
\end{equation} 
where $\phi= \betaI/(\betaI+\betaR)$. The first and second terms of {Eq.~}\eqref{eq:PC} are due to diagnosis during asymptomatic and symptomatic phases, respectively. Recovering without diagnosis happens with probability $\mathcal{P}_\text{R} = 1-\mathcal{P}_\text{C}$. Therefore, we can determine a simple relation between the final number of documented ($N_\text{C}$)  and undocumented ($N_\text{R}$) infections defining the under-reporting coefficient $\sigma_\text{ur}$ as
\begin{equation}
\sigma_\text{ur}=\frac{N_\text{R}}{N_\text{C}}=\frac{1-\mathcal{P}_\text{C}}{\mathcal{P}_\text{C}} = 
\frac{(1-\pA)(1-\phi \pI)}{\pA+(1-\pA)\pI\phi}\,,
\label{eq:sigma}
\end{equation} 
where 
\begin{equation}
N_\text{C}=N_{\text{C}_\text{A}} + N_{\text{C}_\text{I}} + N_{\text{R}_\text{A}} + N_{\text{R}_\text{I}} + N_\text{D}\,.
\label{eq:N_C}
\end{equation}

We can also analytically  determine  the model's  IFR, represented by $\ell_\text{IFR}$, considering the probabilities that exposed individuals evolve to death passing through C$_\text{A}$ compartment or not, which are $\pA\phi \frac{ \eta}{\eta+\alpha_\text{R}}$ or $(1-\pA) \phi \pI \frac{\eta}{\eta+\alpha_\text{R}}$, respectively. The IFR becomes
\begin{equation}
\ell_\text{IFR} = [\pA+(1-\pA)\pI]\frac{\phi\eta}{\eta+\alpha_\text{R}}\,.
\label{eq:ell_IFR}
\end{equation}

\subsection{Estimating under-reporting from epidemic surveillance counts}
\label{sub:ur}

The rates $\muA$, $\betaI$, $\betaR$, $\alphaR$, and $\eta$ are biological and can, in principle, be found in epidemiological surveys~\cite{Davies2020, Li2020b, Nishiura2020, He2020, Baud2020,  Lauer2020}. The parameters $\lbA$ and $\lbI$ depend on behavioral aspects such as the number of potential infectious contacts  per unit of time~\cite{Aleta2020b,Arenas2020,Goyal2021}; prophylactic attitudes by means of NPIs~\cite{Candido2020, Moore2021, DiDomenico2020}; infectiousness and prevalence of new variants~\cite{Castro2021,Davies2021,Kupferschmidt2021}; to cite only some of the most prominent issues. Similarly, the confirmation rates $\betaC$ and $\alphaC$ depend on several behavioral and socioeconomic factors being highly influenced by testing policies~\cite{Ahmed2020,Sun2021,Aleta2020b}. All these aspects are very heterogeneously distributed across time and different places.

We describe how testing probabilities can be estimated from surveillance count series with the aid of the compartmental model of Fig.~\ref{fig:modelo}. Let $\mathcal{C}(t)$ and $\mathcal{D}(t)$ represent the cumulative series of confirmed cases and deaths. The CFR  computed for reported cases  within a given time window  $[t_\text{cal},t_\text{cal}+\Delta \tau]$ is given by:
\begin{equation}
\ell_\text{CFR} \equiv \frac{\Delta \mathcal{D}(t_\text{cal})}{\Delta \mathcal{C}(t_\text{cal}-t_\text{delay})}\,,
\label{eq:ell_CFR_data}
\end{equation}
in which $\Delta\mathcal{C}$ and $\Delta\mathcal{D}$ refer, respectively, to the increment in the number of cases and deaths in the  interval, $t_\text{cal}$ is the initial time chosen for calibration, and $t_\text{delay}$ is a delay between death and positive test report. The CFR is given by the conditional probability that an infection evolves to death given that it was diagnosed. If $\mathbb{D}$ and $\mathbb{T}$ are events of death and diagnosis of infected individual, respectively, we have that
\begin{equation}
\ell_\text{CFR} \equiv \Pr(\mathbb{D}|\mathbb{T}) = \frac{\Pr(\mathbb{D} \cap \mathbb{T})}{\Pr(\mathbb{T})} =  \frac{\Pr(\mathbb{D})}{\Pr(\mathbb{T})} = 
\frac{\ell_\text{IFR}}{\mathcal{P}_\text{C}}\,,
\label{eq:ell_CFR_model}
\end{equation}
where we have used the Bayes rule for conditional probabilities, the model hypothesis that only diagnosed individuals evolve to death, and Eq.~\eqref{eq:PC}. Rearranging the terms, one obtains
\begin{equation}
\frac{\ell_\text{IFR}}{\ell_\text{CFR}} = \pA+(1-\pA)\phi\pI\,.
\label{eq:ellratio}
\end{equation}

Despite its simplicity, Eq.~\eqref{eq:ellratio} is very handy since it relates the testing rates (or probabilities) with epidemiological parameters ($\ell_\text{IFR}$ and $\phi$) and quantities ($\ell_\text{CFR}$) which can, in principle, be obtained directly from data, Eq.~\eqref{eq:ell_CFR_data}. Therefore, if the ratio $r=\pA/\pI$ is given, the testing rates can be estimated from Eq.~\eqref{eq:PC} as 
\begin{equation}
p_\text{I} = \frac{r+\phi}{2r\phi}\left[1-\left(1-\frac{4r\phi \mathcal{P}_\text{C}}{(r+\phi)^2} \right)^{1/2}\right].
\label{eq:pI}
\end{equation}
Equation~\eqref{eq:pI} is mathematically consistent, \textit{i.e.} $0\le p_\text{I}\le 1$, if the following conditions are satisfied
\begin{equation}
\begin{matrix}
\mathcal{P}_\text{C} & \le &  r(1-\phi)+\phi & \text{for} & 0\le r\le \frac{\phi}{2\phi-1}\,, \\
\mathcal{P}_\text{C} & \ge &  r(1-\phi)+\phi & \text{for} &  r\ge \frac{\phi}{2\phi-1}\,. \\
\end{matrix}
\end{equation}
The under-reporting coefficient, which can be expressed as
\begin{equation}
\sigma_\text{ur} = \frac{1-\mathcal{P}_\text{C}}{\mathcal{P}_\text{C}} = \frac{\ell_\text{CFR}}{\ell_\text{IFR}}-1,
\label{eq:sigma_cfr}
\end{equation}
is extracted directly from data using Eq.~\eqref{eq:ell_CFR_data} and is used to estimate $\pA$ and $\pI$ for a given ratio $r$, which may play a role on the determination of infection rates $\lbA$ and $\lbI$; see Sec.~\ref{sub:hidden}. 

A sensibility analysis of the ratio $r$ can be used to verify whether the results are little sensitive to this choice (it was the case for all data analyzed in this work); otherwise the ratio must be determined using some calibration or likelihood method. Eventually, surveillance data can provide a CFR smaller than the IFR estimates which is inconsistent with the {present approach and the method is not applicable in these situations}.

\subsection{Assessing hidden compartments from epidemic surveillance data}
\label{sub:hidden}

Epidemiological surveillance provides the number of confirmed cases, deaths, date of first symptoms, or diagnosis;  nothing with respect to the other compartments is commonly available.  Actually, in the real scenario,  the situation is much more complicated due to delays and other complex issues on surveillance counts~\cite{Bastos2019,Bastos2020}. Using the case series $\mathcal{C}(t)$, we estimate infection rates $\lbA$ and $\lbI$ concomitantly with the initial conditions $({S}^*,{E}^*,{A}^*,{I}^*,{R}^*)$  using the {following} calibration procedure:

\begin{itemize}[leftmargin=1.2em]
	\item[i.]  Select the time interval $[t_\text{cal},t_\text{cal}+\Delta \tau]$ for which the reported case series $\mathcal{C}(t)$ will be analyzed. This time window should be short enough to assume that infection rates $\lbA$ and $\lbI$ are approximately constant, but sufficiently large to have significant amount of data;
	
	\item [ii.] Using the time series of case and death counts, determine the probability $\pI$ using Eq.~\eqref{eq:pI} for a given ratio $r=\pA/\pI$, assumed to be a parameter of the method.
	
	\item[iii.] Consider an adiabatic approximation assuming that susceptible population varies much more slowly than the other compartments such that one can neglect its variation and take $S(t)\approx S^*$ as being constant over the investigated period. 
	
	\item[iv.] Start with guessed initial values for the products  $\gammaA=\lbA S^*$ and $\gammaI=\lbI S^*$ (to be fitted with data).
	
	\item[v.] Determine the number of undocumented cases $N_\text{R}^*$ at $t=t_\text{cal}$ using the under-reporting coefficient calculated using Eqs.~\eqref{eq:sigma}, \eqref{eq:ell_CFR_data}, \eqref{eq:pI}, and \eqref{eq:sigma_cfr} and the number of confirmed cases $N_\text{C}^*=\mathcal{C}(t_\text{cal})-\mathcal{C}(t_\text{tr})$ from case counting, where $t_\text{tr}$ is a transient time to be chosen accordingly the epidemiological series. Remember that $N_\text{C}^*$ encompasses all confirmed compartments; see Eq.~\eqref{eq:N_C}.
	
	\item[vi.] Under these conditions the compartmental model provides a closed linear system  $\dot{\mathbf{X}}=\mathbb{J}\mathbf{X}$ for the infectious compartments
	$\mathbf{X}=(E,A,I)$  where the Jacobian is given by
	\begin{equation}
	\mathbb{J}= \begin{bmatrix}
	-\muA& \gammaA & \gammaI  \\
	\muA & -(\betaI+\betaR+\betaC)& 0 \\
	0 & \betaI&-(\alphaR+\alphaC)  
	\end{bmatrix}\,.
	\end{equation}
	We assume that the solution is ruled by the leading term  $\mathbf{X}\sim \mathbf{v}_1 \exp[\Lambda_1 (t-t_\text{cal})] $ where $\mathbf{v}_1 = (v_\text{E},v_\text{A},v_\text{I})$ is the principal eigenvector corresponding to the largest eigenvalue $\Lambda_1$ of $\mathbb{J}$,  providing the following relations among initial conditions $({E}^*,{A}^*,{I}^*)$
	\begin{align}
	\frac{E^*}{A*}\approx  \frac{v_\text{E}}{v_\text{A}}\, \text{~and~} 	\frac{I^*}{A^*}\approx  \frac{v_\text{I}}{v_\text{A}}\,.
	\end{align}
	Using again $\mathbf{X}\sim \mathbf{v}_1 \exp[\Lambda_1 (t-t_\text{cal})]$, a closed system of initial conditions for $({E}^*,{A}^*,{I}^*)$ is obtained with the integration of Eq.~\eqref{eq:dCdt} to obtain
	\begin{equation}
	\Delta \mathcal{C} \approx (\betaC A^* +\alphaC I^*) \frac{e^{\Lambda_1 \Delta \tau}-1}{\Lambda_1}\,,
	\end{equation}
	where $\Delta \mathcal{C}$ is the increment of confirmed cases, available from data, during the interval $\Delta\tau$. If $\Lambda_1 \Delta \tau\ll 1$ we obtain 
	\begin{equation}
	\betaC A^* +\alphaC I^* \approx \frac{\Delta \mathcal{C}}{\Delta \tau}\,.
	\end{equation}
    Finally, the susceptible population is determined as 
	\begin{equation}
%	N_\text{S}^*=N-\sum_{\text{X}\ne \text{S}} N_\text{X},
    N_\text{S}^* =N-N_\text{C}^*-N_\text{E}^*-N_\text{A}^*-N_\text{I}^*\,,
	\label{eq:N_S}
	\end{equation}
	implying that $S^*=N_\text{S}^*/N$, and the infection rates self-consistently estimated as $\lbA=\gammaA/S^*$ and $\lbI=\gammaI/S^*$.
	
	\item[vii.] Equations~\eqref{eq:dEdt} to \eqref{eq:dCdt} are integrated in the interval $[t_\text{cal},t_\text{cal}+\Delta \tau]$  and the dispersion with respect to the case counts is computed as 
    \begin{equation}
       \Omega(\gammaI,\gammaA) = \int_{t_\text{cal}}^{t_\text{cal}+\Delta \tau} \left[\mathcal{C}(t)-C(t)\right]^2 \text{d}t\,.
    \end{equation}	
	\item[viii.] The parameters $\gammaA$ and $\gammaI$ are incremented interactively and steps (iv) to (vii) are implemented using a bisection method to minimize $\Omega(\gammaI,\gammaA)$. {In other words, a mesh with discrete values of $(\gammaI,\gammaA)$, with mesh space $(\Delta \gammaI,\Delta \gammaA)$, is varied searching for the minimal value of $\Omega(\gammaI,\gammaA)$. Then, the mesh space is reduced and the analysis repeated around the pair $(\gammaI,\gammaA)$ that yielded the lowest $\Omega$ is the preceding step. {This process} is iterated		
		{1000 times}.} 
\end{itemize}
The choice of the transient time $t_\text{tr}$ should compensate new epidemic factors such as new variants and waning immunity that lead to reinfections and new outbreaks. Another factor that can alter the susceptible population is the vaccination which also confers {variable} levels of immunity against infections. Vaccination also impacts both the IFR and CFR, such that the updated estimates of the IFR should be considered if the count series fueling the analysis is concomitant with vaccination, as the case of our current analysis; see Sec.~\ref{sub:parameters}.

\section{Results}

\label{sec:results}

\subsection{Parameters and epidemic series}
\label{sub:parameters}

We applied the method to two types of count series available for Brazil, hereafter named Type-I and Type-II. The former consists of count series using  release dates provided by epidemic surveillance departments of Brazilian federative units\footnote{Brazil is divided into 26 states and 1 federal district. States aggregate municipalities with independent administrative structure. Cases are reported by municipalities to state's healthcare departments which release the information publicly.} which are aggregated and publicly available for all 5570 Brazillian municipalities~\cite{wcota}. These data do not yield the date of diagnosis  and may present uncontrolled bias caused by reporting delays and should be used with care. The Type-II data sets contain dates of diagnosis and first symptoms onset. In this work, we use the publicly available Type-II data for Paraná (PR)~\cite{covid19PR} and Espirito Santo (ES)~\cite{covid19ES} states. The data are publicly available in the cited resources and the data aggregated for different municipalities, used in the present work, is available elsewhere~\cite{codes}. A full description of these datasets can be found in the Supplementary Material (SM)~\cite{SM}. 

We fixed the average values of the parameters $\muA^{-1}=3.2$~d and $\betaI^{-1}=3.2$~d so that the mean incubation time is of 6.4~d~\cite{Li2020b,Arenas2020}.  The mean recovery time for symptomatic individuals was taken as $\alphaR^{-1}=3.2$~d~\cite{Read2021}. Following~\cite{Arenas2020,Costa2020}, asymptomatic cases were assumed to have the same recovering time such that $\betaR^{-1}=\betaI^{-1}+\alphaR^{-1}$. Uncertainty analysis was done drawing $\muA$, $\betaI$, and $\alphaR$ from Gamma distributions with standard deviation of $1.3$~d, while an uniform distribution with 10\% of uncertainty were used for calibrated $\gammaA$ and $\gammaI$.

\begin{figure}[th]
	\centering
	\includegraphics[width=0.9\linewidth]{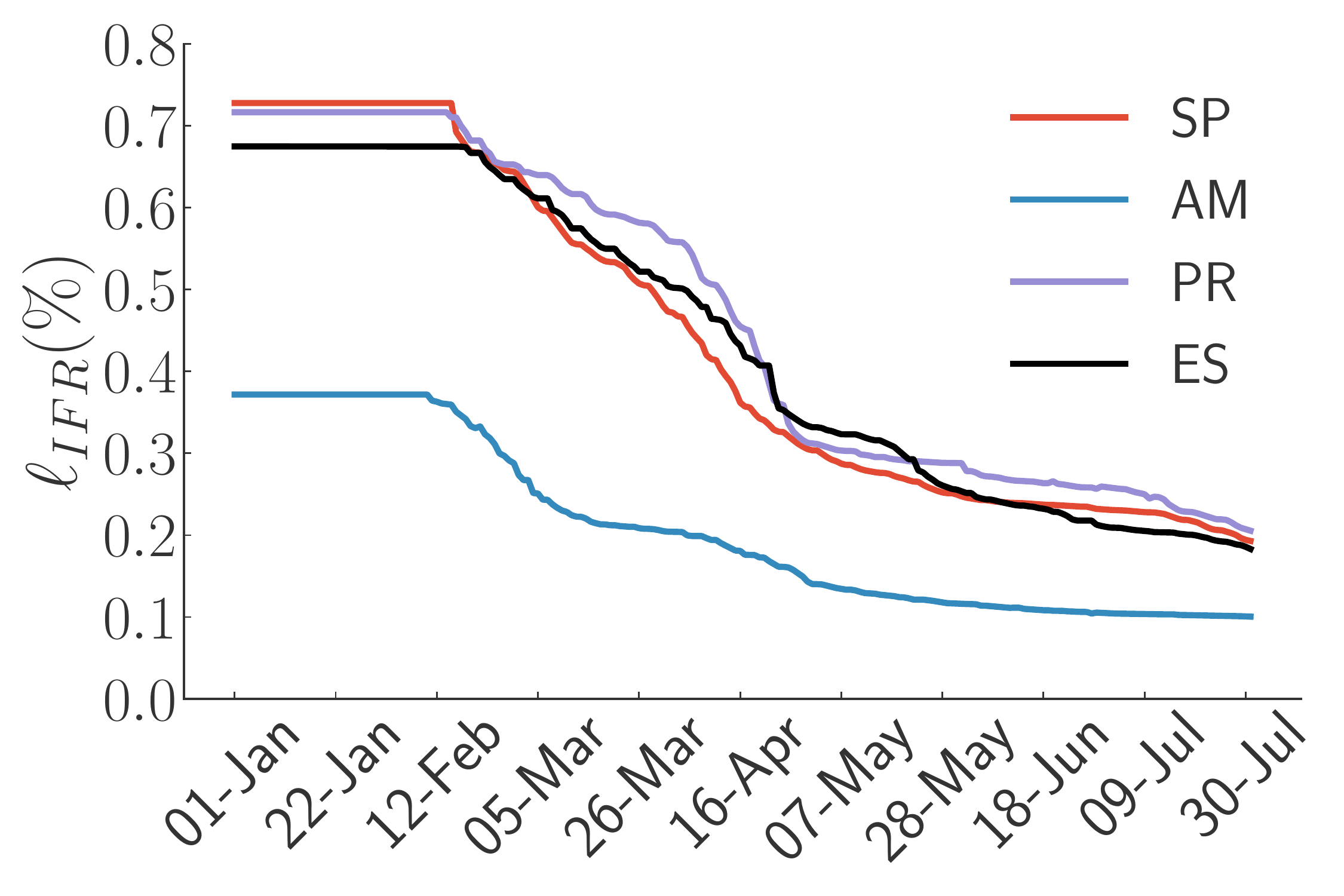}
	\caption{Infection fatality ratio as a function of time, estimated for  São Paulo (SP), Amazonas (AM), Paraná (PR), and Espírito Santo (ES) states considering their demographics and vaccination rates.}
	\label{fig:ifr}
\end{figure}
The IFR is the most critical parameter of our analysis. Since the time window we analyzed is concomitant with vaccination, a progressive reduction in the IFR is expected. To estimate the IFR reduction due to vaccines we proceeded as follows. The age-dependent IFR profile reported by Verdity~\cite{Verity2020}, which yields an exponential increase with age and average IFR 0.68\%, was considered.  The number of  persons who completed the vaccination (two or one shot depending on the vaccine type) as a function of time was extracted from surveillance systems, publicly available at Ref.~\cite{wcota}. Demographic data were obtained from \textit{Instituto Brasileiro de Geografia e Estatística} (IBGE)~\cite{instituto2017divisao}.  Brazil followed a decreasing age prioritization strategy where elderly were vaccinated first down to the young population. We consider $g=1,\ldots,N_\text{g}$ group ages, in which $g=1$ corresponds to $\ge 75$~yr, $g=2$ to $70 - 74$~yr,\ldots, $g=16$ to $0-4$ and assume that all vaccines shots were distributed according to this sequence. Using data for states, both vaccination rates and demographics~\cite{instituto2017divisao}, we calculated the average IFR as follows. Without vaccines, the average IFR is given by
\begin{equation}
\ell_\text{IFR} = \sum_{g=1}^{N_\text{g}} \ell_g n_g,
\end{equation}
where $\ell_g$ and $n_g$ are, respectively, the IFR and population fraction in the age group $g$. If $x$ is the total fraction of the vaccinated population, the lower age  group $g^*$ who were vaccinated is given by
\begin{equation}
\sum_{g=1}^{g^*} f_g n_g<x<\sum_{g=1}^{g^*+1} f_g n_g,
\end{equation}
where $f_g$ is  the fraction of the group age $g$ who was vaccinated. Finally, if $r_g$ is the IFR reduction of the vaccinated population of age group $g$, the corrected {IFR} becomes
\begin{equation}
\ell_\text{IFR} = \sum_{g=1}^{N_\text{g}} \ell_g n_g-\sum_{g=1}^{g^*} \ell_g n_g f_g(1-r_g) \,.
\end{equation}
For sake of simplicity, we assumed that $f_g=f=0.85$ and $r_g=r=0.05$ uniform across all age groups. These parameters are consistent with typical protection rates associated to vaccines used in Brazil (Pfizer-Biotech, Sinovac and Astrazeneca). The IFR as a function of time for the four investigated states are presented in  Fig.~\ref{fig:ifr}. The lower IFR for Amazonas's state reflects its young population (see SM~\cite{SM}), while similar patterns are observed for the other analyzed states. Obviously, this is a simplified approach aiming at being qualitatively correct rather than quantitatively accurate. The used data is available in the SM~\cite{SM}.

\subsection{Under-reporting coefficient}
\begin{figure}[!t]
	\centering
	\includegraphics[width=0.98\linewidth]{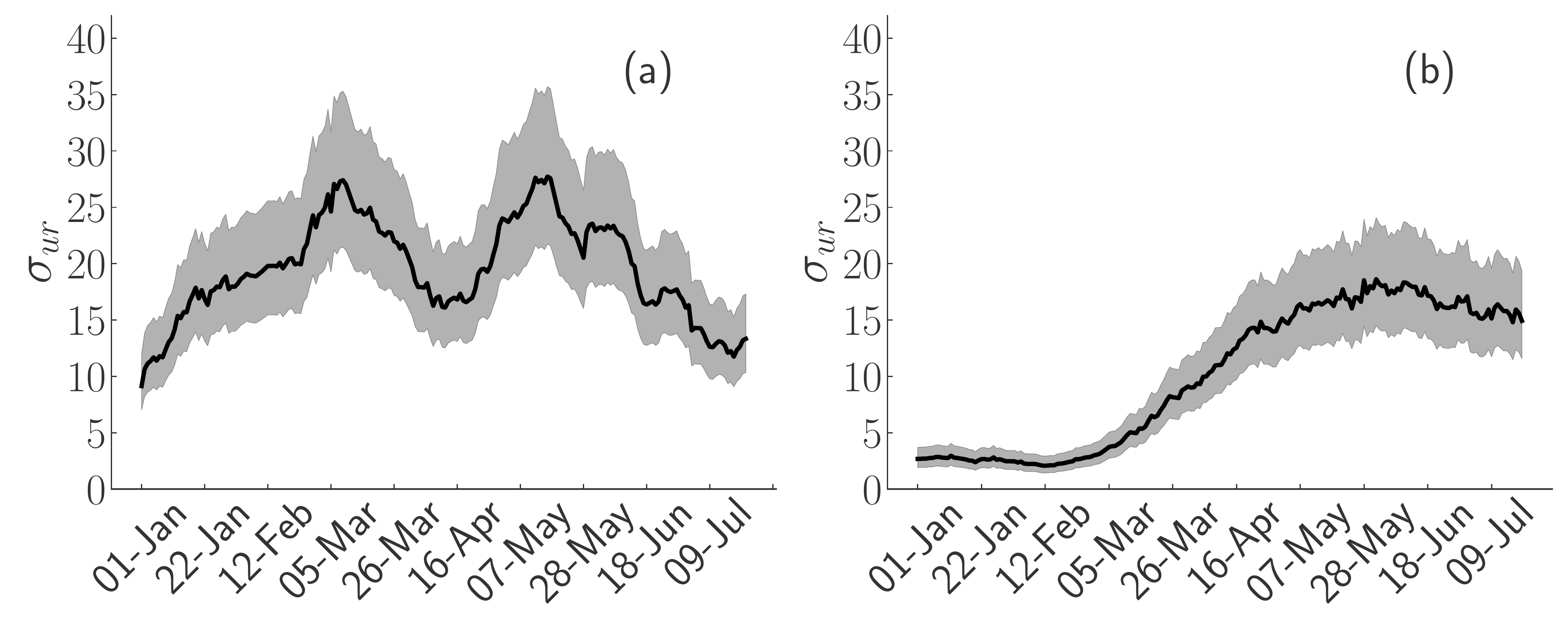}
	\caption{Evolution of under-reporting coefficient $\sigma_\text{ur}$ for the capital cities of (a) Manaus and (b) São Paulo estimated using moving time windows of three weeks for Type-I count series (see main text) as reported by state's surveillance departments~\cite{wcota}. The {confidence interval} of 95\% is shown in the shaded region. 
	}
	\label{fig:underrep}
\end{figure}
\begin{figure*}[t]
	\centering
	\includegraphics[width=0.3\linewidth]{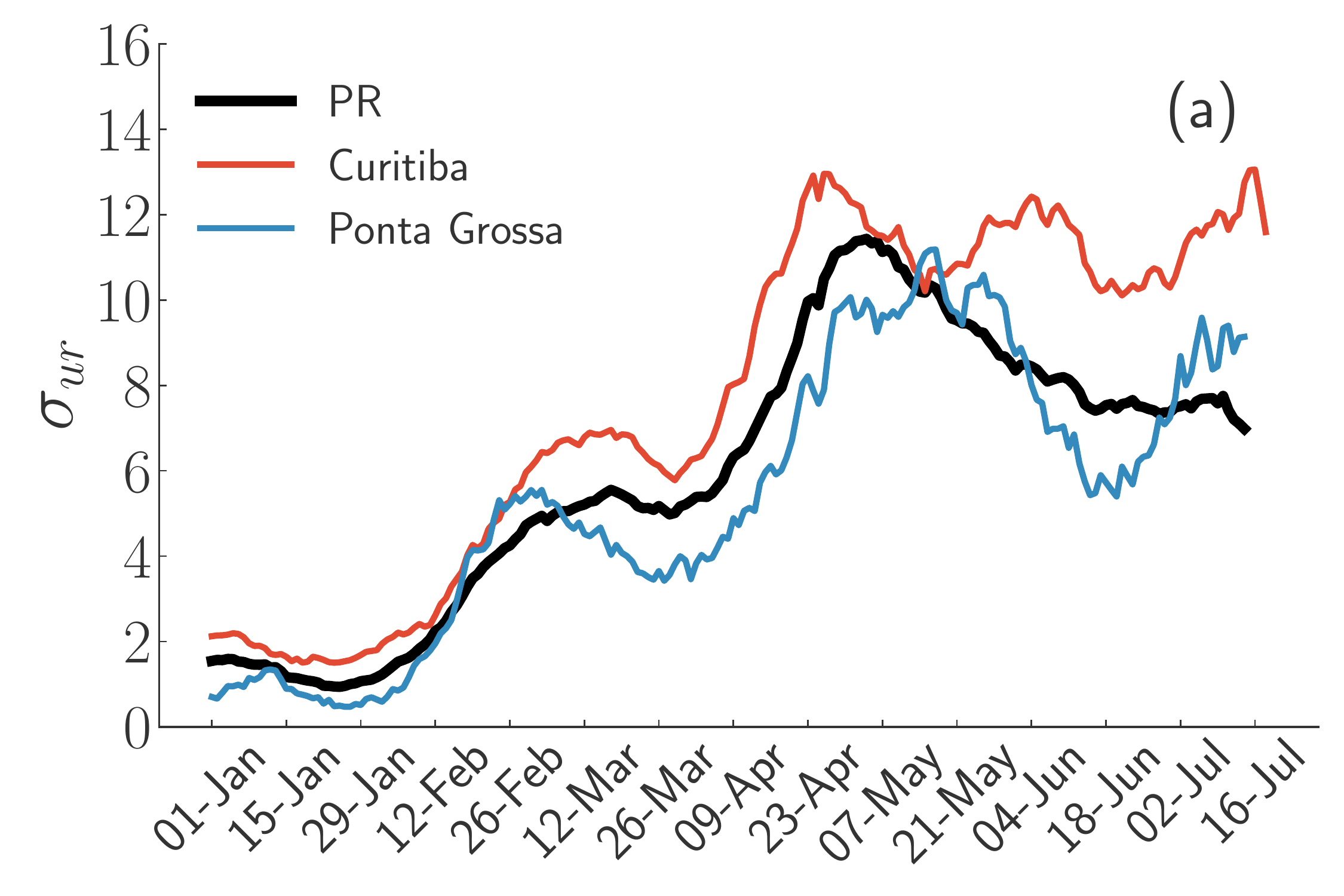}  
	\includegraphics[width=0.3\linewidth]{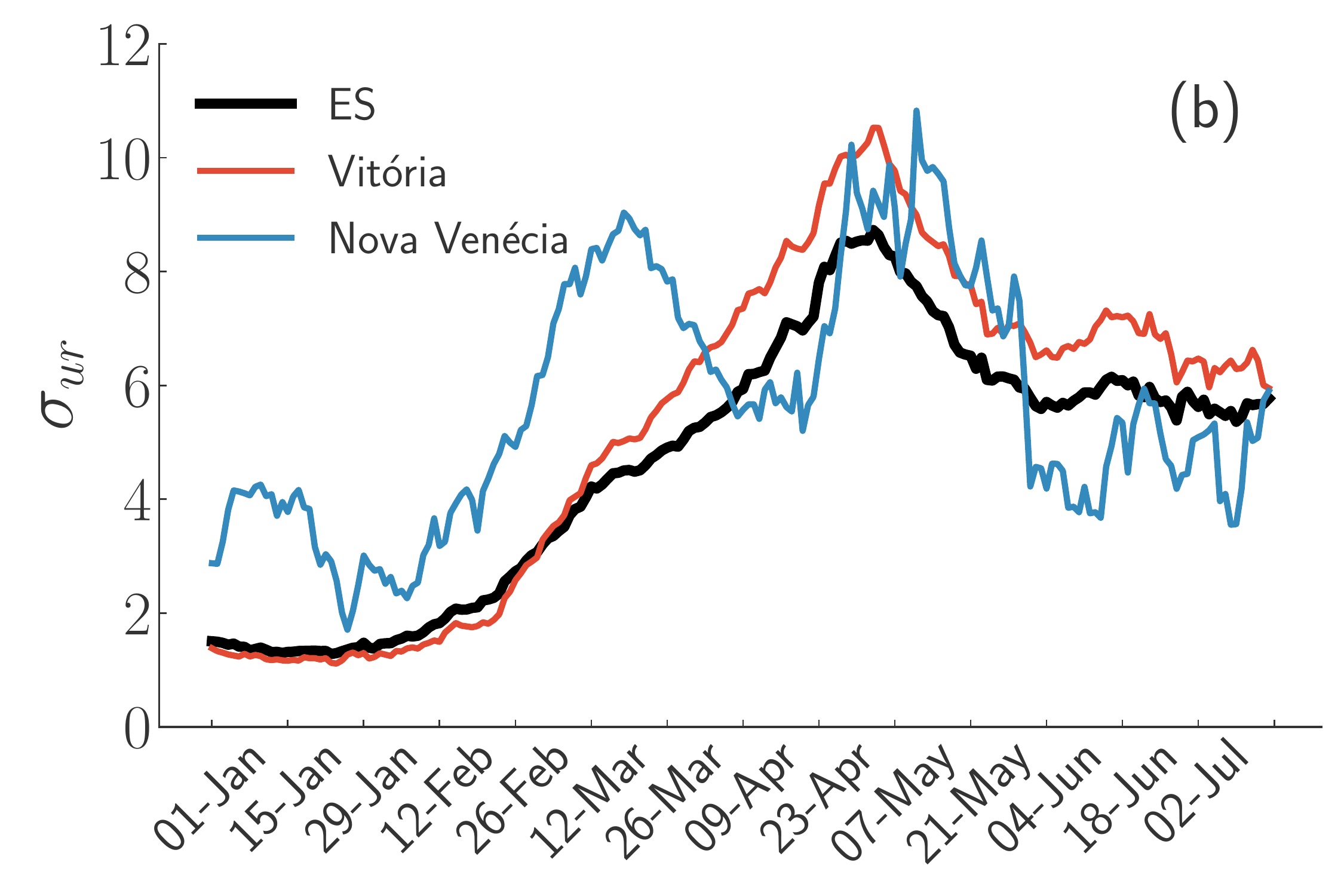} 
	\includegraphics[width=0.3\linewidth]{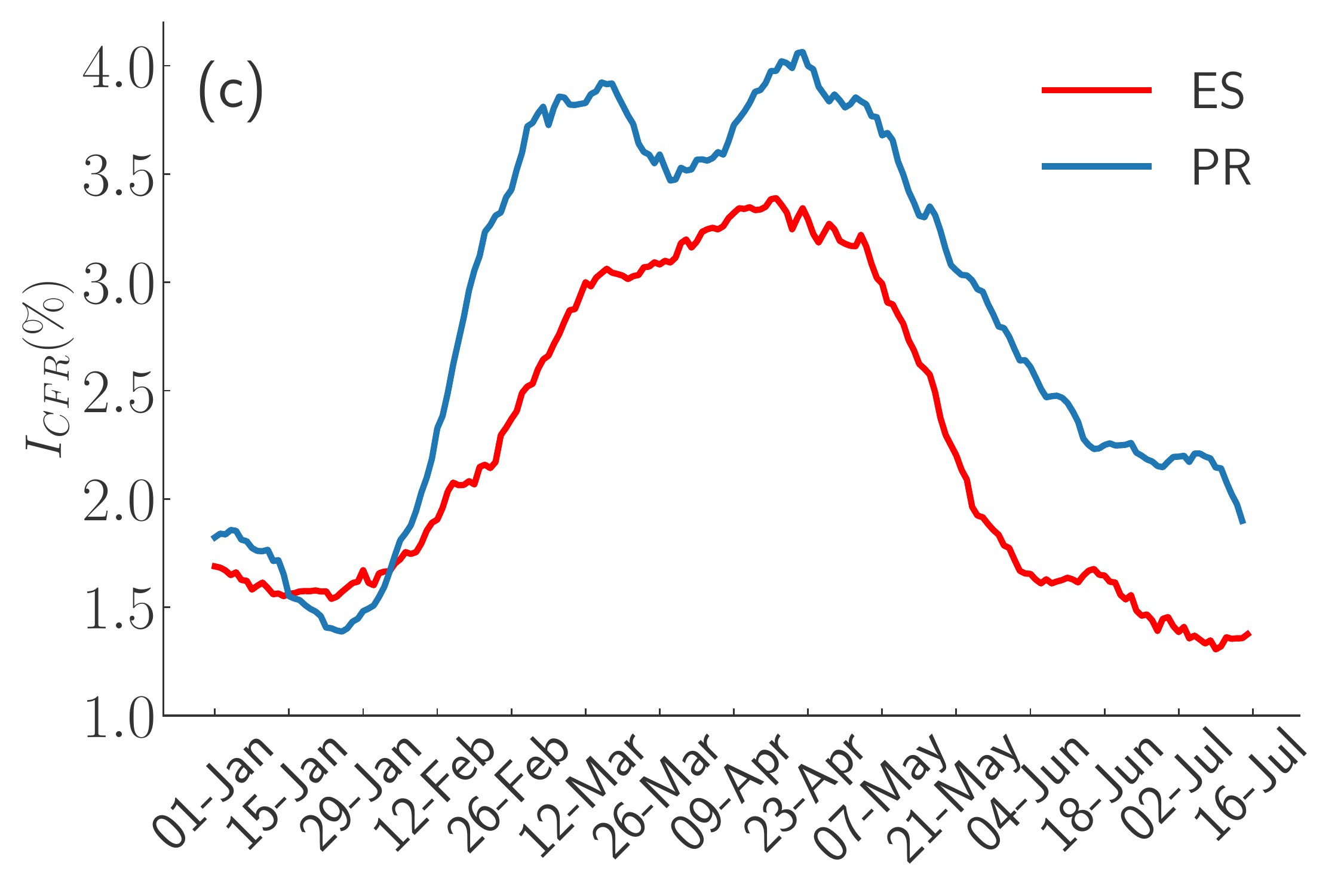}\\
	\includegraphics[width=0.45\linewidth]{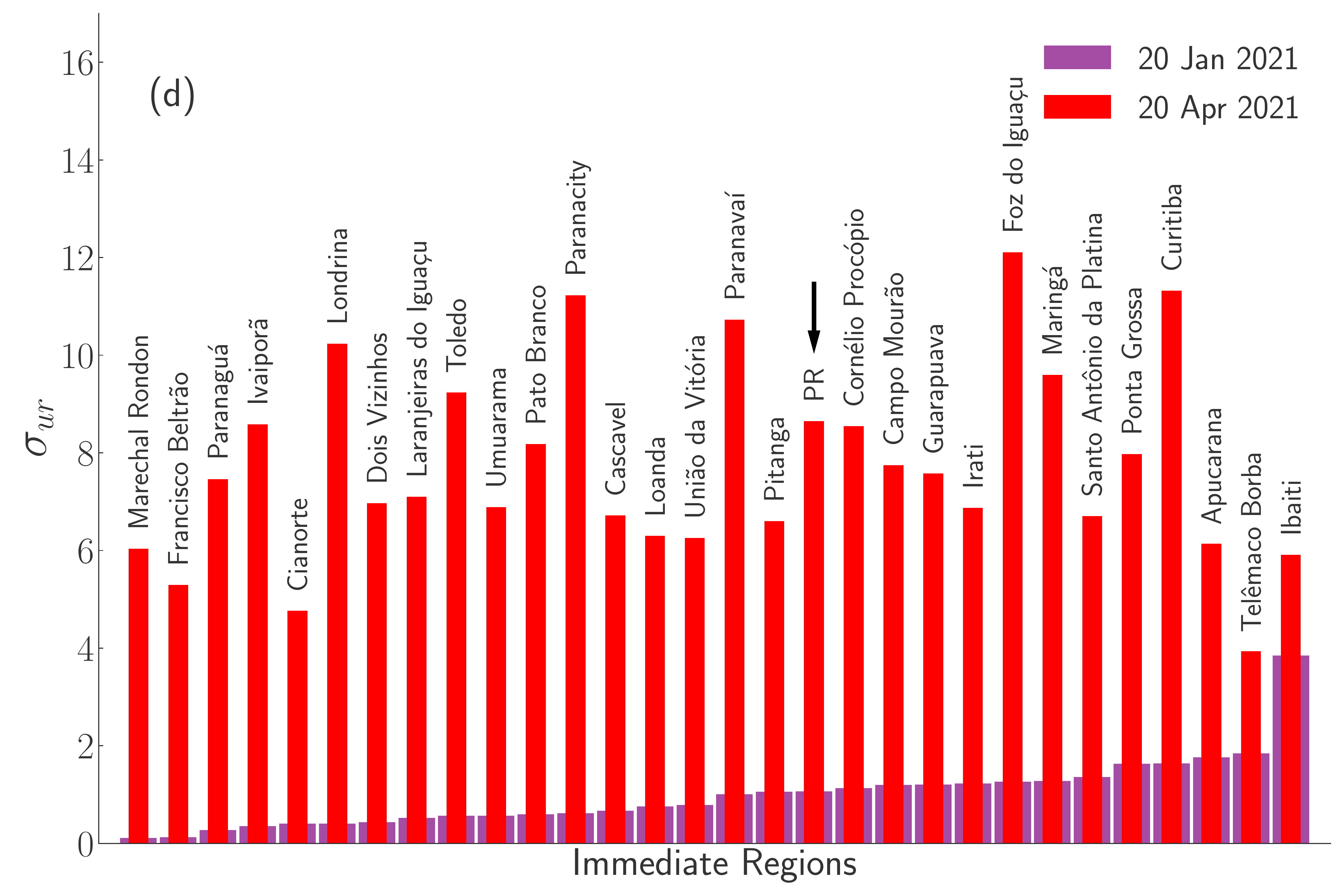} 	
	\includegraphics[width=0.42\linewidth]{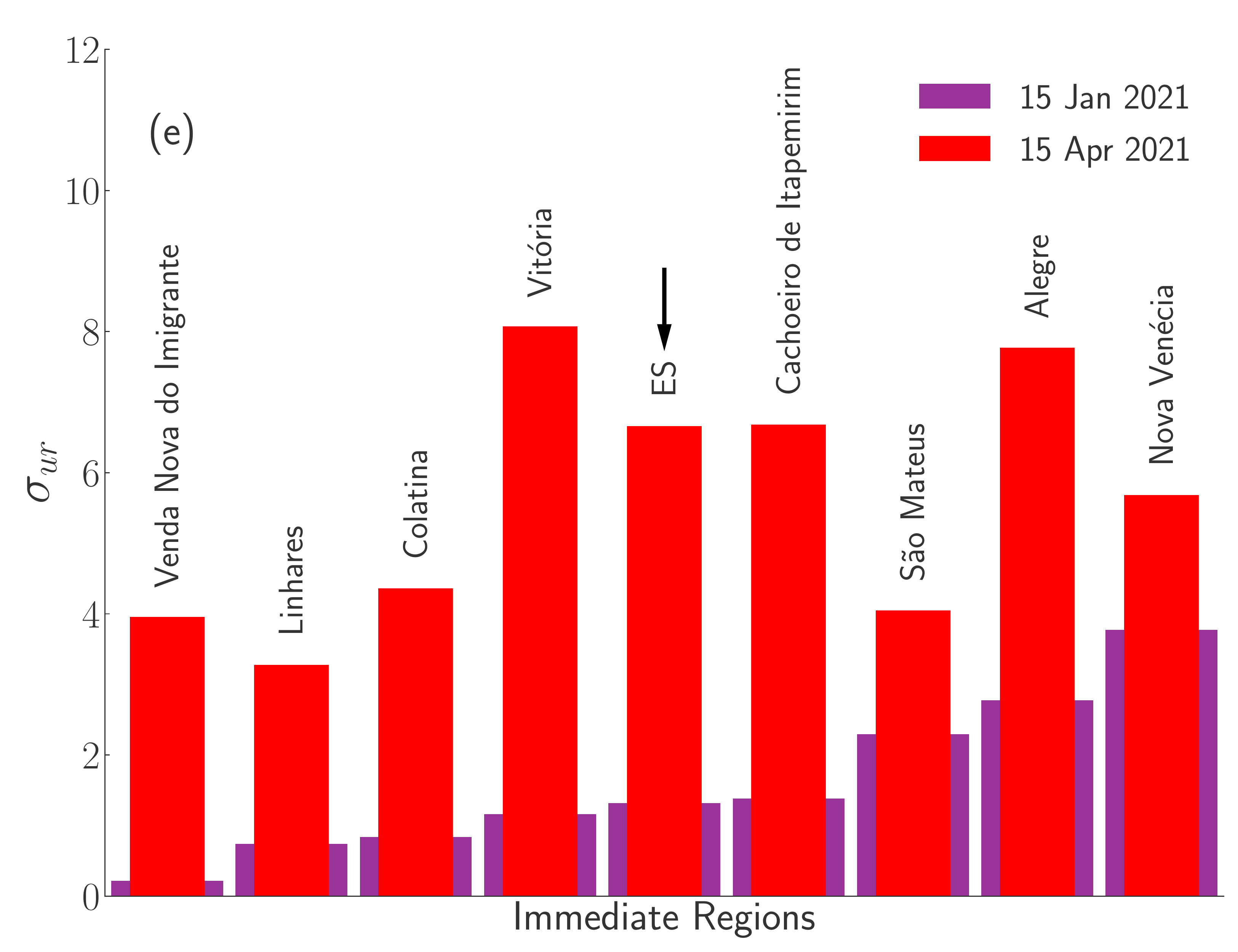}
	\caption{Evolution of under-reporting coefficient for (a) PR and (b) ES states using  time windows of three weeks. Two immediate regions of each state are presented in the corresponding panels. (c) Evolution of the CFR computed using delays $t_\text{delay}=10$~d and 20~d for PR and ES states, respectively. Under-reporting coefficients for all immediate regions of (d) PR and (e) ES and the for states (indicated by arrows) computed when the CFR is low (January 2021) and high (April 2021).}
	\label{fig:sigmaimedpr}
\end{figure*}

The evolution of  $\sigma_\text{ur}$ using  Type-I count series of two capital cities of Brazil, which were severely impacted by COVID-19 second infection wave, namely Manaus and São Paulo~\cite{Castro2021}, are presented in Fig.~\ref{fig:underrep}, for which the estimated delays between case and death confirmations were $t_\text{delay}=7$ and $9$ days,  respectively; see Figs. S1(a) and (b)in the SM~\cite{SM}. The delay is obtained by shifting the time series such that the peaks of deaths and cases coincide. We consider $t_\text{tr}$ as January 1, 2021. Evolution patterns of $\sigma_\text{ur}$ are  different for these municipalities. While Manaus presents a high level of under-reporting  (10 to 25) along the whole analyzed time series, in São Paulo, $\sigma_\text{ur}$ increases from approximately 3 at the beginning of 2021 to 17 in June.  

We analyzed Type-II count series for PR and ES states  aggregating data of municipalities  into immediate regions defined by IBGE~\cite{instituto2017divisao} as a group of nearby municipalities of a same state with intense interchange for immediate needs (purchasing, work, healthcare, education, and so on). Case and death series for the PR state present a delay of $t_\text{delay}\approx 10$~d between death and positive test report.  For ES state this delay {is} $t_\text{delay}\approx 20$~d. 

The evolution of $\sigma_\text{ur}$ computed with counts aggregated by states and two selected immediate regions are shown in Figures~\ref{fig:sigmaimedpr}(a) and (b) for PR and ES, respectively. Curves for the 28 and 8 immediate regions of PR and ES, respectively, with the confidence intervals are available in Figs. S2 and S3 of the SM~\cite{SM}. Note that CFR , Fig.~\ref{fig:underrep}(c), and $\sigma_\text{ur}$ present different temporal patterns despite the correlation stated by Eq.~\eqref{eq:sigma_cfr}. The second relevant outcome  is  the substantial variation of undocumented infection along the time and across different places with $\sigma_\text{ur}$ varying approximately one order of magnitude in Figs.~\ref{fig:sigmaimedpr}(a) and (b). The under-reporting coefficient for all immediate regions of both PR ans ES states are presented in Figs.~\ref{fig:sigmaimedpr}(d) and (e); the chosen dates correspond to low and high CFR in the respective state counts. The differences between immediate regions can {differ largely} in a same time window. The space-time variability reflects the high  diversities of outbreak across different places, due to  unsynchronized and unequal responses to pandemics besides demographic, economic, and developmental heterogeneity of states as predicted~\cite{Costa2020} and later observed~\cite{Castro2021} for the first epidemic wave in Brazil.

\subsection{Determination of the  initial conditions}
\begin{figure*}[th]
	\centering
	\includegraphics[width=0.7\linewidth]{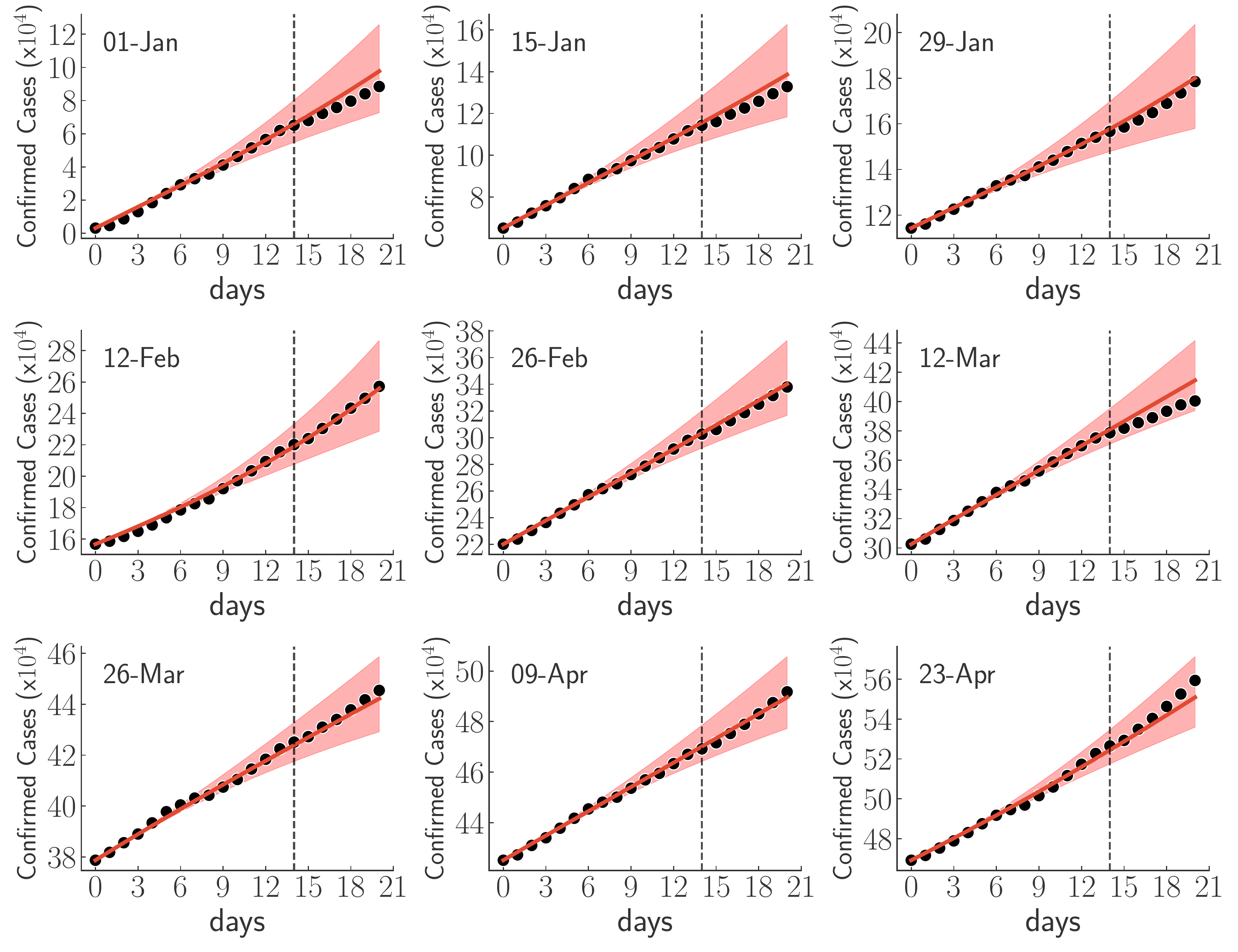} 
	\caption{Calibration curves  for PR state in different time windows of 14 days indicated by the vertical lines.  Initial day is {indicated} in the top of each panel. One week of forecasting is also shown. Symbols are the cumulative cases' counts  while lines with shaded regions represent the calibrated curves and the corresponding confidence interval of 95\%.}
	\label{fig:prcalibra}
\end{figure*}

\begin{figure}[th]
	\centering
	\includegraphics[width=0.9\linewidth]{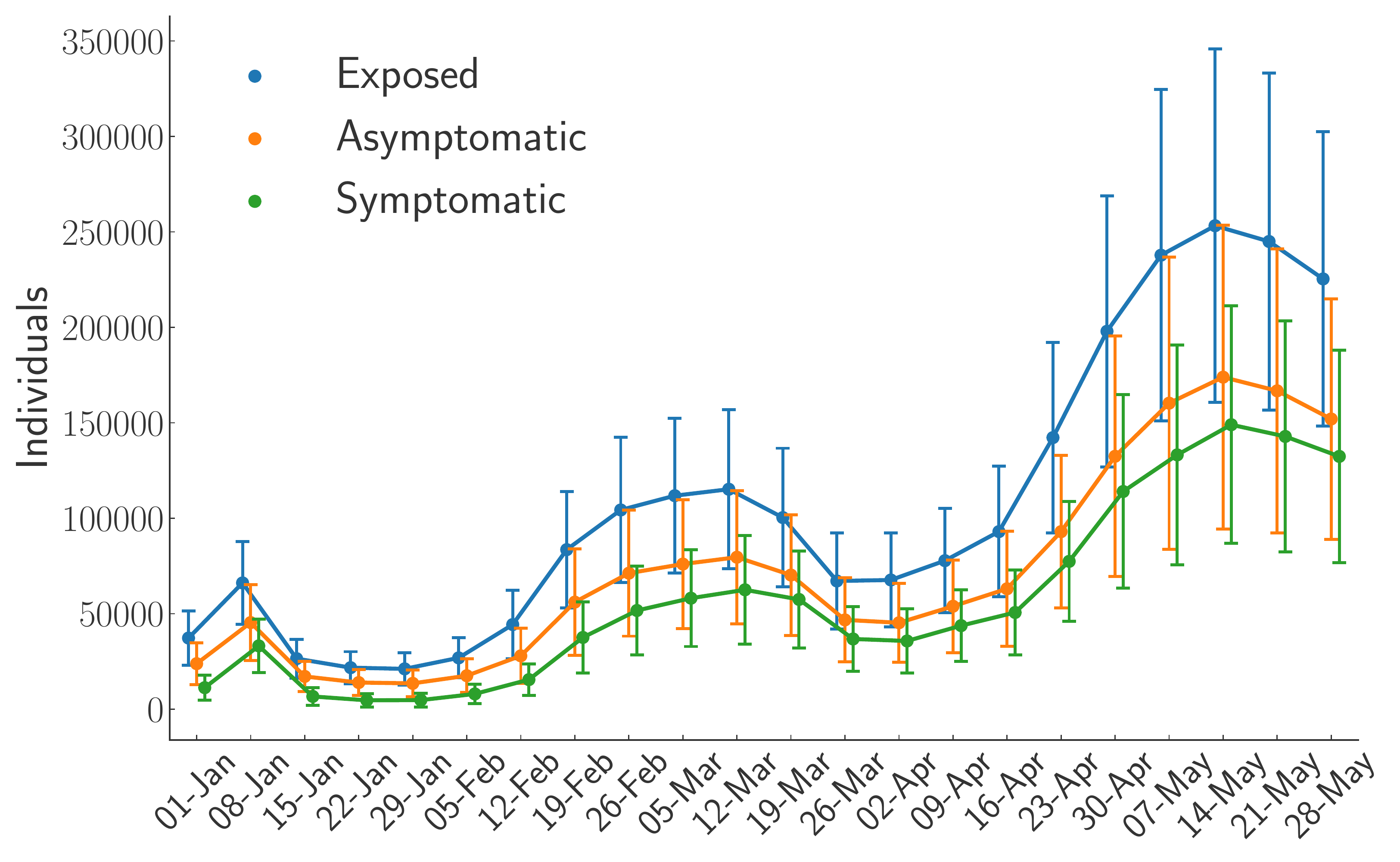}
	\caption{Evolution of the undocumented compartments (exposed, asymptomatic and symptomatic) for the PR state since January 1, 2021.}
	\label{fig:prEAU}
\end{figure}

To apply the calibration method of Sec.~\ref{sub:hidden}, we performed the analysis  for case counts of the PR state shown in Fig.~\ref{fig:prcalibra}; see Fig. S4 on SM~\cite{SM} for the ES state. We further simplified the analysis assuming the same infection rate for both asymptomatic and symptomatic individuals prior diagnosis confirmation, $\lambda_\text{A}=\lambda_\text{I}$, implying in a single parameter to fit the data. {The ratio between testing probabilities of symptomatic and asymptomatic individuals is fixed to $\pA/\pI=0.1$. The calibrated curves match each other within the confidence intervals  for a variation of one order of magnitude in this ratio.} Typical calibration curves are presented in Fig.~\ref{fig:prcalibra}(a)-(i) for different times using a 14-day moving window of calibration. A forecast of one week is also presented to verify the calibration robustness, reproducing very well the short-term progression of the cumulative case count time series. The method also performs  very well for smaller geographical scales such as immediate regions; see Fig. S5 of the SM~\cite{SM}. 

The evolution of the undocumented epidemic compartments (exposed, asymptomatic, and symptomatic) yielded by the calibration method  for PR state from January to May 2021 is presented in Fig.~\ref{fig:prEAU}. Remark that the ratio between the total amount of infected individuals and the number of confirmed cases at a given day is much higher than the under-reporting coefficient shown in Fig.~\ref{fig:sigmaimedpr} since the latter refers to the final epidemic chain, where an infection ends documented or not, whereas the former refers to the amount of infected individuals in a given day which has not been documented yet. The peaks of prevalence of infectious cases happen slightly before peaks of incidence of confirmed cases. {Figure~\ref{fig:prEAU} shows an increase of the unconfirmed cases in the same period (middle April to May of 2022) when the under-reporting was higher for the PR state; Fig.~\ref{fig:underrep}(a). One explanation for this behavior is the vaccination which leads to less aggressive manifestation of the infections and lower seeking for medical attention and testing.}

\begin{figure}[ht]
	\centering
	\includegraphics[width=0.9\linewidth]{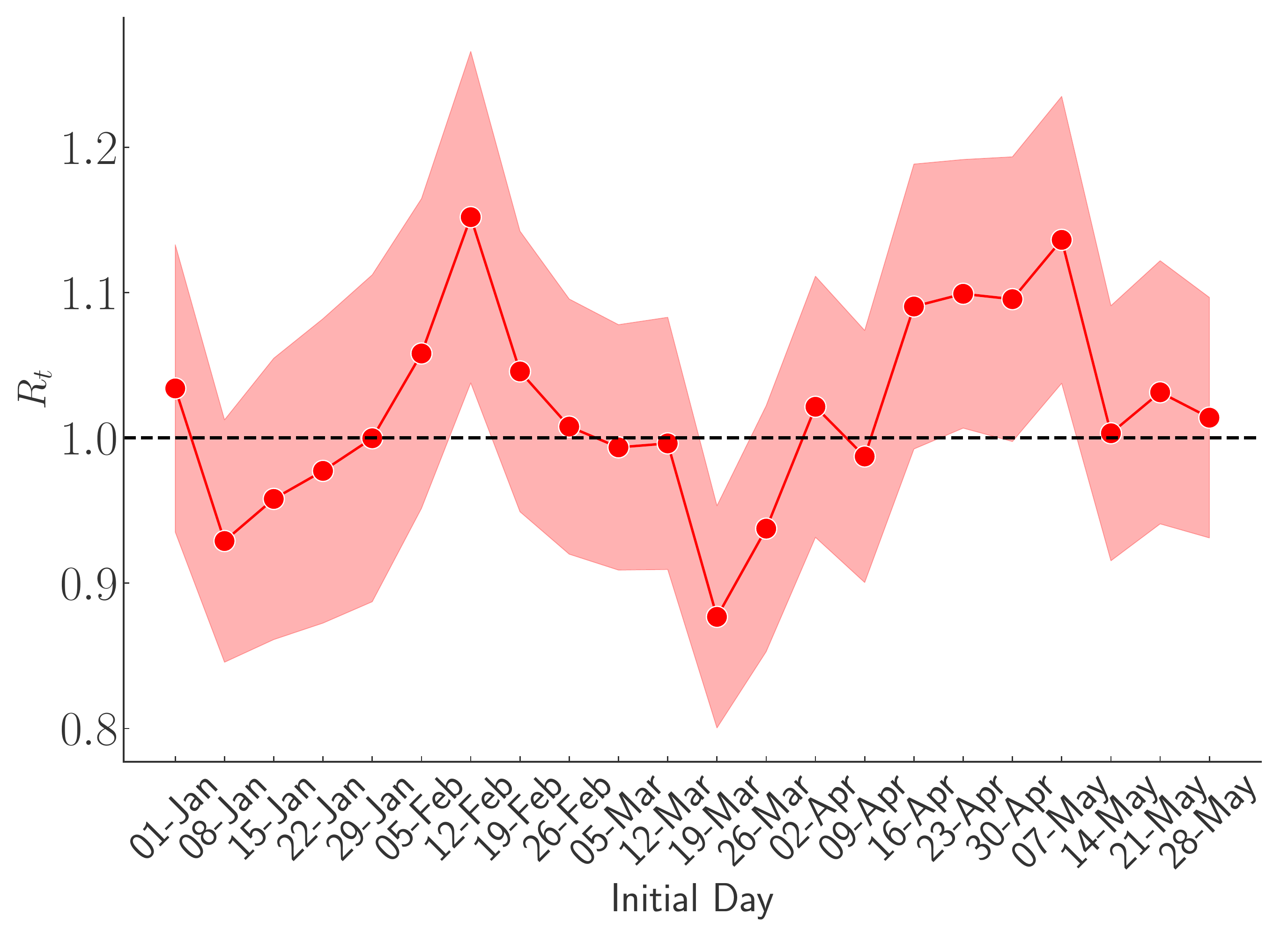}
	\caption{Evolution of effective reproductive number $R_t$ computed for the PR state considering $p_A/p_I = 0.1$ and $\lbA = \lbI$. The confidence interval of 95\% is shown in the shaded region for the black curve.}
	\label{fig:rtpr}
\end{figure}

The effective reproduction number for PR state is presented in Fig.~\ref{fig:rtpr}. The calibration is sensitive to the variations and inflections in case count series, where the mean value of $R_t$ oscillated between approximately 0.9 and 1.2. We performed a sensibility analysis  of $R_t$ and verified that its value is almost independent of the testing rates of asymptomatic compartments. More precisely, the curves of $R_t$ collapses within the confidence interval when the ratios between testing probabilities $p_\text{A}/p_\text{I}$ and infection rates $\lambda_\text{A}/\lambda_\text{I}$ are varied by one order of magnitude. 

\section{Discussion}
\label{sec:conclu}

The pandemic caused by the SARS-CoV-2 led to unprecedented efforts gathering scientific community, epidemic surveillance, public authorities, and communication systems to provide  almost real-time updated and publicly available counts for diagnosed infections, deaths, and  other important statistics for COVID-19 spread across the globe. Available epidemic series, however, are still not ideal due to our limited capacity in documenting all infections in the due time. Moreover, these limitations vary enormously across different places and {at different} moments. However, this opens new avenues for construction and improvement of tools to extract information which are not explicit in data. A particularly promising strategy is the data-driven approach~\cite{Aleta2020,Arenas2020,Costa2020} where mathematical and mechanistic models are fueled by data, allowing to estimate variables which are not explicitly available. In the case of {SARS-CoV-2 infections}, the important class of asymptomatic or pre-symptomatic infections, in which individuals transmit the pathogen even without symptoms, are crucial being very costly to be detected in epidemic surveillance systems.

In the present work, we follow a data-driven  approach using a  compartmental model to estimate the amount of undocumented cases in the epidemic compartments which are not directly accessible in surveillance systems. The method allows to estimate the fraction of undocumented infections using case fatality ratio (CFR) and biological parameters that, in principle, can be estimated in controlled studies, in particular the infection fatality ratio (IFR). We applied the method to epidemic series of diagnosed cases and deaths of two Brazilian states where days of the symptoms onset were available. We selected the first semester of 2021 when Brazil was struck by a second epidemic wave of COVID-19, mainly driven the Gamma variant (lineage P.1). We calculated a under-reporting coefficient $\sigma_\text{ur}$, giving the ratio between infections which ends diagnosed or not. Our analysis reports a large variation of $\sigma_\text{ur}$ along the time and also across different locations at a same period. The method allows to estimate the initial condition for the undocumented compartments, in particular the asymptomatic and exposed ones.  While, on the one hand, the presented numbers should be not interpreted as accurate estimates of actual epidemic prevalence, on the other hand, they clearly demonstrate that the infected individuals that  can potentially seek for medical assistance are a minor part of all cases. Interestingly, the effective reproduction number is almost insensitive to the testing rate of asymptomatic cases, confirming that undocumented infections do not affect this important epidemic indicator.

The method can be generalized for stratified data including age contact matrices~\cite{Prem2017} or metapopulation approaches~\cite{Arenas2020,Costa2020}. However, the main lesson is that initial conditions for undocumented compartments can be inferred using a simple mechanistic approach, based on compartmental models fueled by {epidemiological} series of  diagnosed death and cases. Nonetheless, the accuracy of methods depends on good estimates of biological parameters, mainly the IFR that changes as the epidemic scenario is altered. For example, vaccination is expected to reduce IFR while the emergence of more aggressive  variants can increase it. We developed a simple data-driven approach to estimate the IFR evolution in terms of {time series with} vaccination rates.  As a forthcoming continuation of the present work, we could investigate different  time distributions for epidemic transitions, akin to applied epidemiology,  using, for example, Monte Carlo approaches.

\bigskip

\noindent  \textbf{Code and data:} Fortran and Python codes used for calibration and processing the epidemic series were made publicly in \cite{codes}. A description of the datasets and codes can be found in the SM~\cite{SM}.

\bigskip

\noindent \textbf{Author contributions}: \textbf{GSC}: Conceptualization; Formal analysis; Data curation; Investigation; Methodology; Software;  Visualization;  Writing - review \& editing.  \textbf{WC}: Conceptualization; Data curation; Methodology; Software; Validation; Visualization; Writing - review \& editing. \textbf{SCF}: Conceptualization;  Formal analysis; Funding acquisition; Methodology; Project administration; Writing - original draft.

\bigskip

\noindent  \textbf{Competing interests:} Authors declare no competing interest.

\section*{Acknowledgments}
{This work was partially supported by the Brazilian agencies \textit{Coordenação
de Aperfeiçoamento de Pessoal de Nível Superior} - CAPES (Grant no. 88887.507046/2020-00),\textit{Conselho Nacional de Desenvolvimento Científico e Tecnológico}- CNPq 		(Grants no. 430768/2018-4 and 311183/2019-0) and \textit{Fundação de Amparo à Pesquisa do Estado de Minas Gerais} - FAPEMIG (Grant no. APQ-02393-18).} This
study was financed in part by the \textit{Coordenação de Aperfeiçoamento de 		Pessoal de Nível Superior} (CAPES) - Brasil  - Finance Code 001.

\end{document}